\documentclass[a4paper]{jpconf}

\pdfoutput=1
 
\usepackage{graphicx}

\begin{document}

\title{Metrology with Atom Interferometry: Inertial Sensors from Laboratory to
  Field Applications}

\author{B Fang, I Dutta, P Gillot, D Savoie, J Lautier, B Cheng, C L Garrido
  Alzar, R Geiger, S Merlet, F Pereira Dos Santos and A Landragin}

\address{LNE-SYRTE, Observatoire de Paris, PSL Research University, CNRS,
  Sorbonne Universit\'es, UPMC Univ.\ Paris 06, 61 avenue de l'Observatoire,
  75014 Paris, France}

\ead{bess.fang@obspm.fr}

\begin{abstract}
Developments in atom interferometry have led to atomic inertial sensors with
extremely high sensitivity.  Their performances are for the moment limited by
the ground vibrations, the impact of which is exacerbated by the sequential
operation, resulting in aliasing and dead time.  We discuss several
experiments performed at LNE-SYRTE in order to reduce these problems and
achieve the intrinsic limit of atomic inertial sensors.  These techniques have
resulted in transportable and high-performance instruments that participate in
gravity measurements, and pave the way to applications in inertial navigation.
\end{abstract}

\section{Introduction}

Since the advent of atom interferometers (AI) in 1990s
(\cite{RiehleOptical1991, KasevichAtomic1991}, see also
\cite{BermanAtom1997} and references therein), tremendous progress has
been made to improve the performance as well as the transportability
of inertial sensors based on AI.  Today, these sensors have already
been used in field applications such as geophysics
measurements~\cite{FarahUnderground2014,HauthAtom2014}, and promise to
surpass other state-of-the-art technologies in inertial
navigation~\cite{GeigerDetecting2011}.  Moreover, they are the prime
candidates to explore some aspects of fundamental physics such as
testing of Einstein Equivalence Principle~\cite{AltschulQuantum2015},
and detection of gravitational
waves~\cite{DimopoulosAtomic2008,GeigerMatterWave2015}.

The performance of AI-based inertial sensors is fundamentally limited by the
noise in the measurement of the transition probability, i.e.\ the standard
quantum projection noise (QPN)~\cite{YverLeducReaching2003}, or ideally the
Heisenberg limit, between the two interferometric states.  Vibration noise on
the ground is however several orders of magnitude higher. The sequential
operation of typical sensors exacerbates the problem by down aliasing
higher-frequency noise, similar to the Dick effect in cold atomic
clocks~\cite{SantarelliFrequency1998}.  Dead time, during which the atoms are prepared or detected, results in loss of information of the inertial
signal, preventing fast averaging of the vibration noise.

Different strategies have been proposed to overcome these difficulties.  When
a (nearly) stable quantity such as the gravitational acceleration $\vec{g}$ is
measured in a stable environment (e.g.\ in a laboratory), one may average the
noise out with longer measurements or perform active
filtering~\cite{PetersHighPrecision2001,HuDemonstration2013}.  When
transportability is required for measurements at different sites, the bulky
active platform must be replaced by passive isolation systems and classical
sensors that measure the aliasing effect as proposed
in~\cite{YverLeducReaching2003} and first demonstrated
in~\cite{LeGouetLimits2008}.  The need to completely eliminate dead time comes
from applications in geophysics and inertial navigation, where the inertial
signals would fluctuate in time and the environment is far from
stable~\cite{JekeliNavigation2005}.

In this paper, we discuss several experiments performed in the
LNE-SYRTE inertial sensor group to address the issues related to the
vibration noise, aliasing and the dead time in order to
explore the intrinsic limits of AI-based inertial sensors.

\section{Inertial Sensors based on Atom Interferometry}
\label{sec:AI}

We start with a brief description of common techniques used in cold-atom
inertial sensors using alkali atoms and stimulated Raman transitions.  These
techniques are by no means unique, but are often practiced for their
technological maturity.  They have been proven reliable in high-performance
instruments. However, the methods for reduction of vibration that we will
describe in the subsequent sections can be extended to other schemes of atom
interferometers.

\begin{figure}[htb!]
\begin{center}
\includegraphics[height=5cm, trim = .5cm .2cm .3cm .6cm]{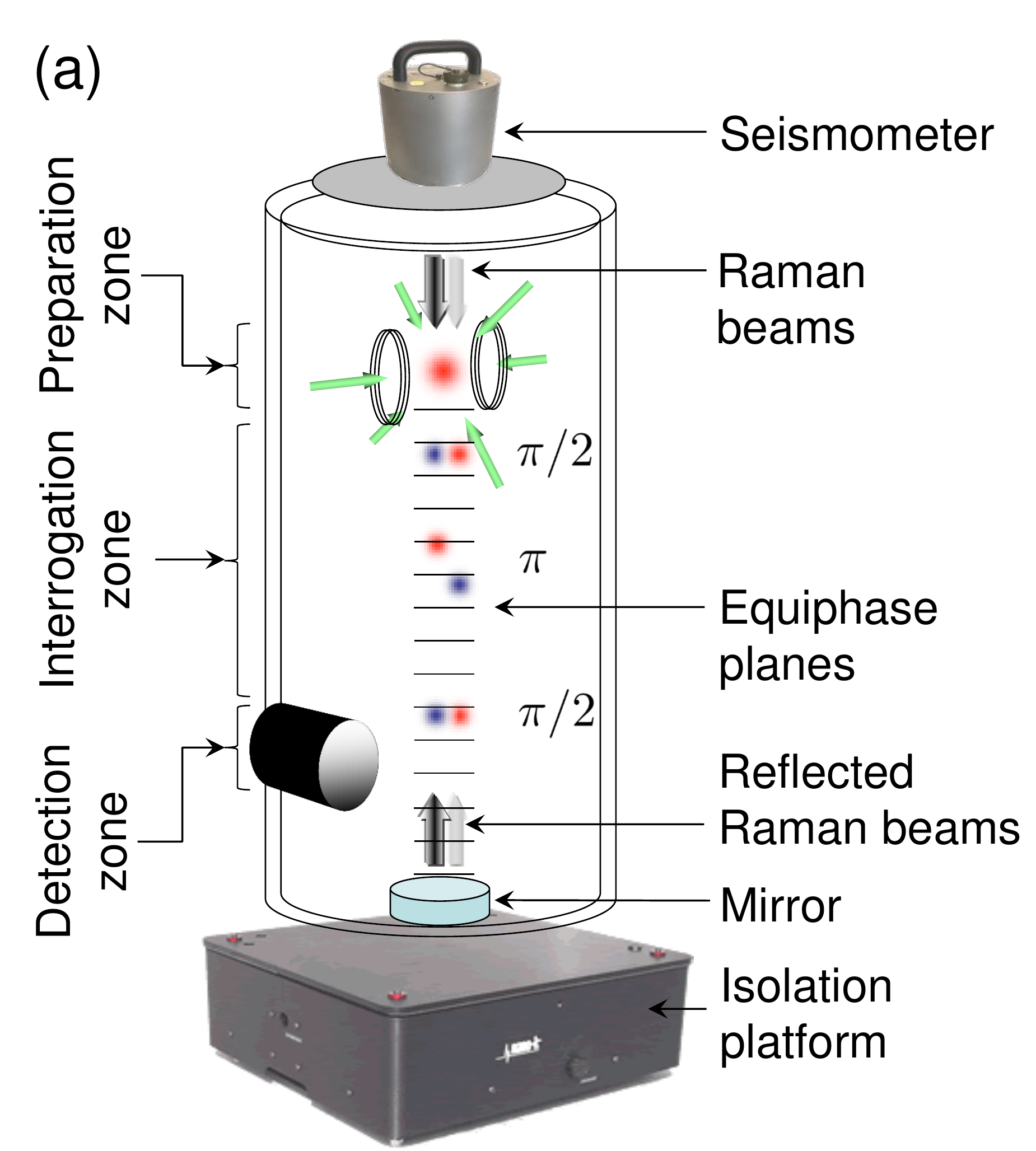}%
\includegraphics[height=5cm, trim = .5cm .55cm .4cm .5cm, clip]{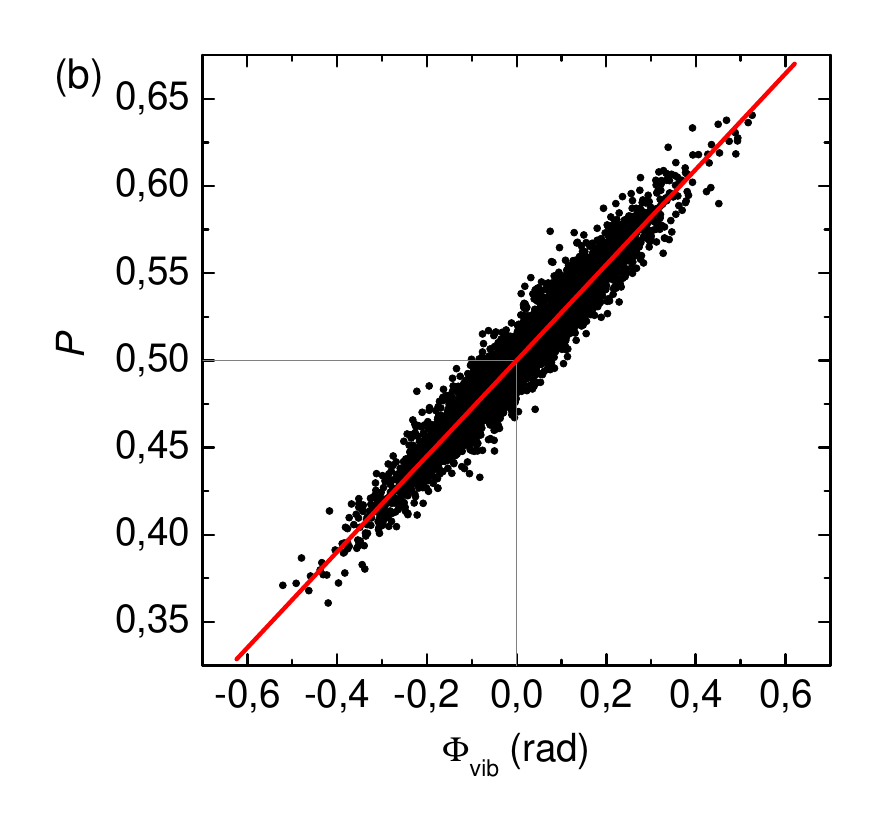}\hspace{.2cm}%
\includegraphics[height=5cm, trim = .5cm .45cm .4cm .5cm, clip]{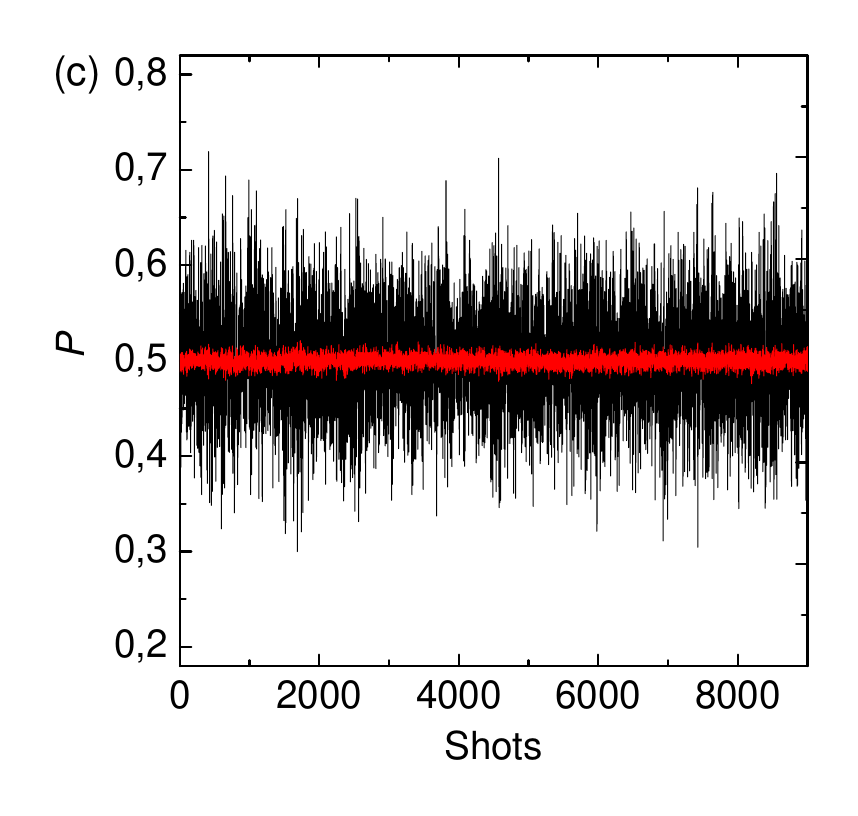}
\end{center}
\caption{(a) Illustration of an atom gravimeter.  It is mounted on a vibration
  isolation platform for passive isolation.  A seismometer is rigidly fixed on
  the structure in order to record residual vibration noise of the
  retroreflection mirror.  (b) Measured transition probability $P$ vs the
  calculated vibration phase $\Phi_{\rm vib}$ (points).  The linear fit
  (straight line) prompts a simple correction procedure by subtraction.  (c)
  Transition probability $P$ of a series of data points before (black) and
  after (red) correlation.  \label{fig:CAG}}
\end{figure}

Before the interferometric measurements start, an atomic source is loaded and
laser cooled to reach sub-Doppler temperatures.  These atoms are then launched
or released into a free fall and enter the interferometric zone.  A
three-pulse atom interferometer is analogous to an optical Mach-Zehnder
interferometer: the matter wave is split, deflected and recombined by
counter-propagating Raman beams via two-photon transitions between two
hyperfine states~\cite{KasevichAtomic1991}.  At the output of the
interferometer, the population in each state is detected by fluorescence.
This gives the transition probability $P$, which depends on the phase
difference $\Delta \Phi$ experienced by the two arms of the interferometer, $P
= \frac{1}{2} (1+ C \cos \Delta \Phi)$, where $C$ is the contrast.  Such a
setup is illustrated in Figure \ref{fig:CAG} (a).

The interferometer phase $\Delta \Phi$ is the difference of the atomic phase
accumulated along the two paths of the interferometer.  It is given by $\Delta
\Phi = \phi_1 - 2\phi_2 + \phi_3$, where $\phi_i$ denotes the phase difference
of the Raman lasers at the center of the atomic wave packet during the $i$th
interrogation pulse.  Since the Raman beams are often retroreflected, the
equiphase planes are defined by the mirror position, making the interferometer
sensitive to the displacement in the direction of propagation of the
beams~\cite{AntoineQuantum2003}.  We can thus write the interferometer phase
as the sum of the desired signal and the noise (e.g.\ vibrations), $\Delta
\Phi = \Phi_{\rm sig} + \Phi_{\rm noise}$.

\section{Vibration Isolation and Correlation with Classical Sensors}
\label{sec:correlation}

We demonstrate here that passively isolating the atomic sensor from the
environment and performing post correlations with classical sensors allow us
to significantly improve the performance of an atom gravimeter and
that of an atom gyroscope.

The LNE-SYRTE Cold Atom Gravimeter (CAG) uses free falling rubidium-87 atoms
in a three-pulse ($\pi/2$-$\pi$-$\pi/2$) configuration.  With the Raman beams
in the vertical direction, the interferometer is sensitive to the
gravitational acceleration $\vec{g}$, i.e.\ $\Phi_{\rm sig} = -\vec{k}_{\rm
  eff} \cdot \vec{g} T^2$, where $\vec{k}_{\rm eff}$ denotes the difference
between the wave vectors of the two Raman lasers, and $T$ is the time interval
between successive Raman pulses.

We operate CAG with a cycling frequency of $3$~Hz and a total interrogation
time $2T = 160$~ms~\cite{LouchetChauvetInfluence2011}.  With about $10^6$
atoms detected, the ground vibration introduces a noise $10^4$ times the
standard QPN limit.  We passively isolate CAG with an anti-acoustic box and a
vibration isolation platform, reducing the vibration noise by about a factor
$100$.  We then acquire the vibration signal of the retroreflection mirror
using a seismometer rigidly fixed on the the frame of CAG [see
  Fig.\ \ref{fig:CAG}~(a)].  The fluctuating velocity of the mirror during the
interrogation results in a shift of the interferometric phase $\Phi_{\rm
  noise} = \Phi_{\rm vib}$.  This can be computed by integrating the acquired
velocity with a weighting function that describes the response of the
interferometer, known as the time transfer function
(see~\cite{CheinetMeasurement2008} and references therein).  Figure
\ref{fig:CAG} (b) shows $P$ versus $\Phi_{\rm vib}$.  The linear relationship
between these two quantities permits a simple correction procedure by
subtraction~\cite{LeGouetLimits2008}.  The vibration rejection efficiency is
about $10$ in the present case [see Fig.\ \ref{fig:CAG}~(c)], and has been
measured to be up to $120$ without passive isolation.

\begin{figure}[htb!]
\begin{center}
\includegraphics[height=6cm, trim = 1.1cm .85cm .8cm .85cm, clip]{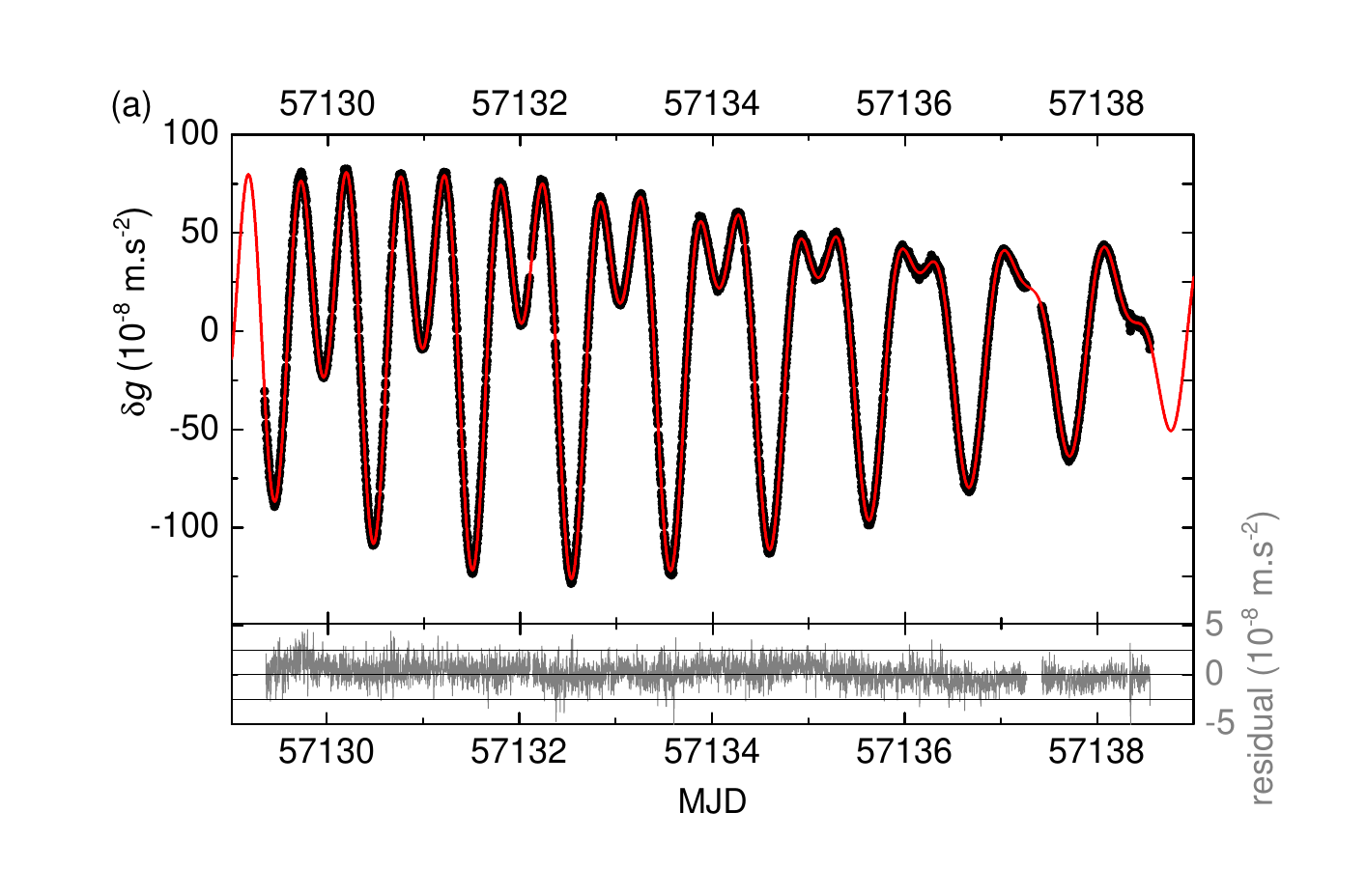}%
\includegraphics[height=6cm, trim = .55cm .5cm .5cm .5cm, clip]{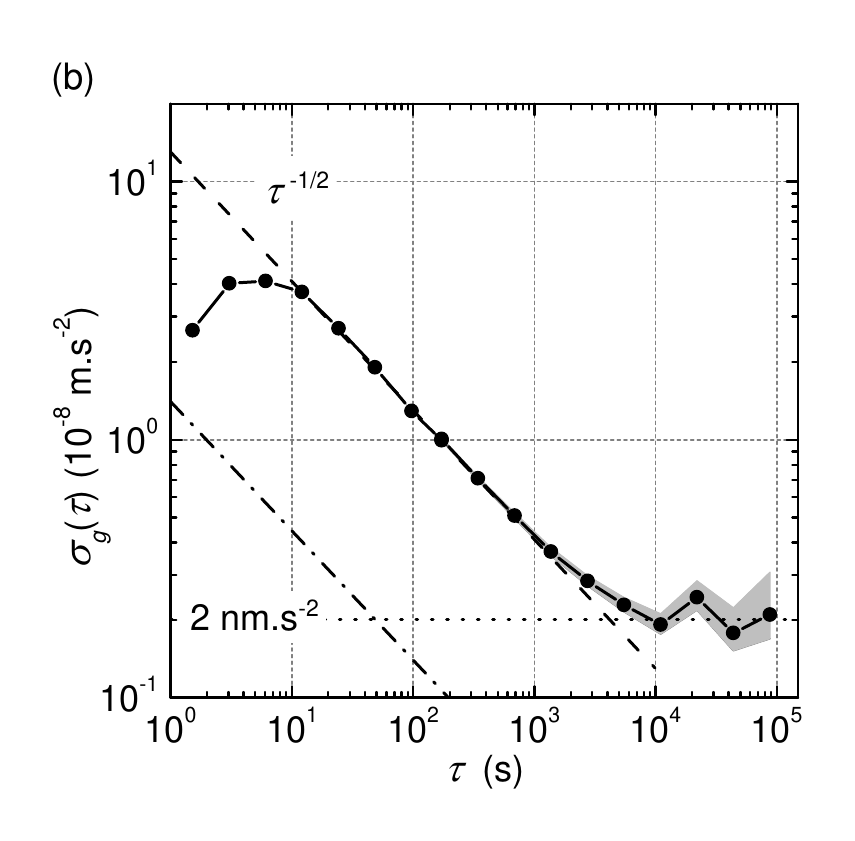}
\end{center}
\caption{\label{fig:CAGcomp} (a) Variation of $g$ around its mean
  measured value as a function of days.  Top: CAG measurement (points)
  averaged over $170$~s and the tide model (line).  Bottom: the
  residual (difference) between the measurement and the model.  (b)
  ADEV of the CAG measurements corrected for the tides and local
  environmental effects.  The shaded region indicates the uncertainty
  of the measurement, whereas the straight line gives the
  $\tau^{-1/2}$ scaling characteristic of averaging white frequency
  noise, where $\tau$ is the integration time.  Similar trend lines
  will be given in subsequent figures with no further explanations.
  The dash-dotted line indicates the calculated QPN limit for typical
  parameters of $10^6$ atoms, $40\%$ contrast, and $T=80$~ms.  }
\end{figure}

We use CAG to record the temporal variation of gravity in our
laboratory dedicated to gravity
measurements~\cite{MerletMicroGravity2008}.  Figure \ref{fig:CAGcomp}
(a) overlays the measured $g$ and the tide model, showing an excellent
agreement throughout the $\pm 10^{-7}$ relative variations around the
mean value due to the lunisolar tide.  The Allan standard deviation
(ADEV) of the measurement corrected for the tides and local
environmental effects is shown in Fig.\ \ref{fig:CAGcomp} (b),
indicating a short-term stability of $13 \times 10^{-8}$~m.s$^{-2}$ at
$1$~s and a long-term stability of $2$~nm.s$^{-2}$ in $3$ hours.  This
is about $10$ times above the calculated QPN limit (dash-dotted line),
as a result of the uncorrelated vibration noise.  In a vibrationally
more stable environment, a best short-term stability of $5.7 \times
10^{-8}$~m.s$^{-2}$ at $1$~s was
demonstrated~\cite{GillotStability2014}.  The accuracy of CAG is
$4.3\times 10^{-8}$~m.s$^{-2}$~\cite{Francis2012}, mainly limited by
the bias caused by the wave-front aberration of the Raman lasers.

\begin{figure}[htb!]
\begin{center}
\includegraphics[height=4cm, trim = 2.5cm 9.1cm 3.5cm 9.7cm, clip]{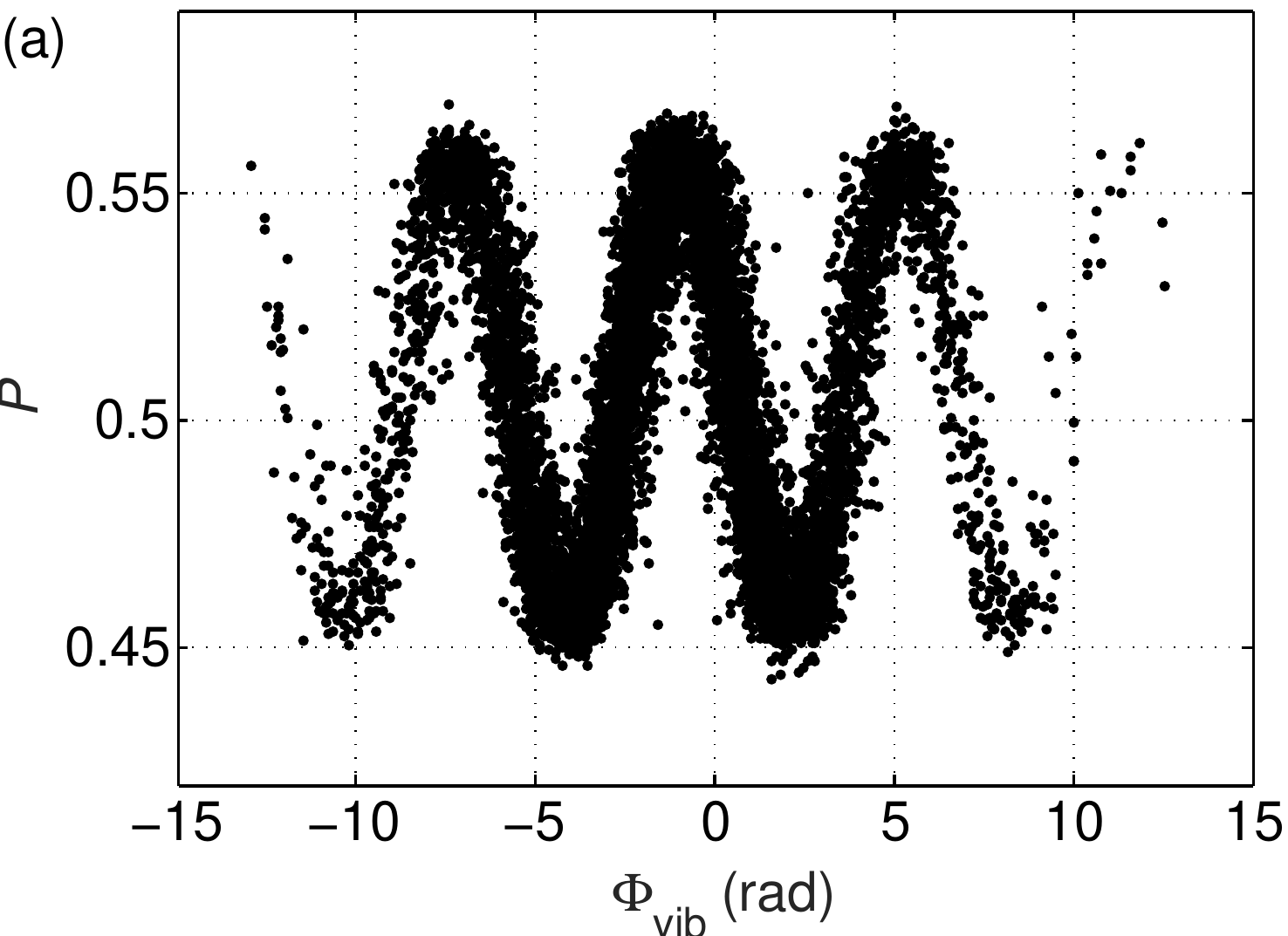}\hspace{.5cm}%
\includegraphics[height=4cm, trim = .6cm .6cm .5cm 1.1cm, clip]{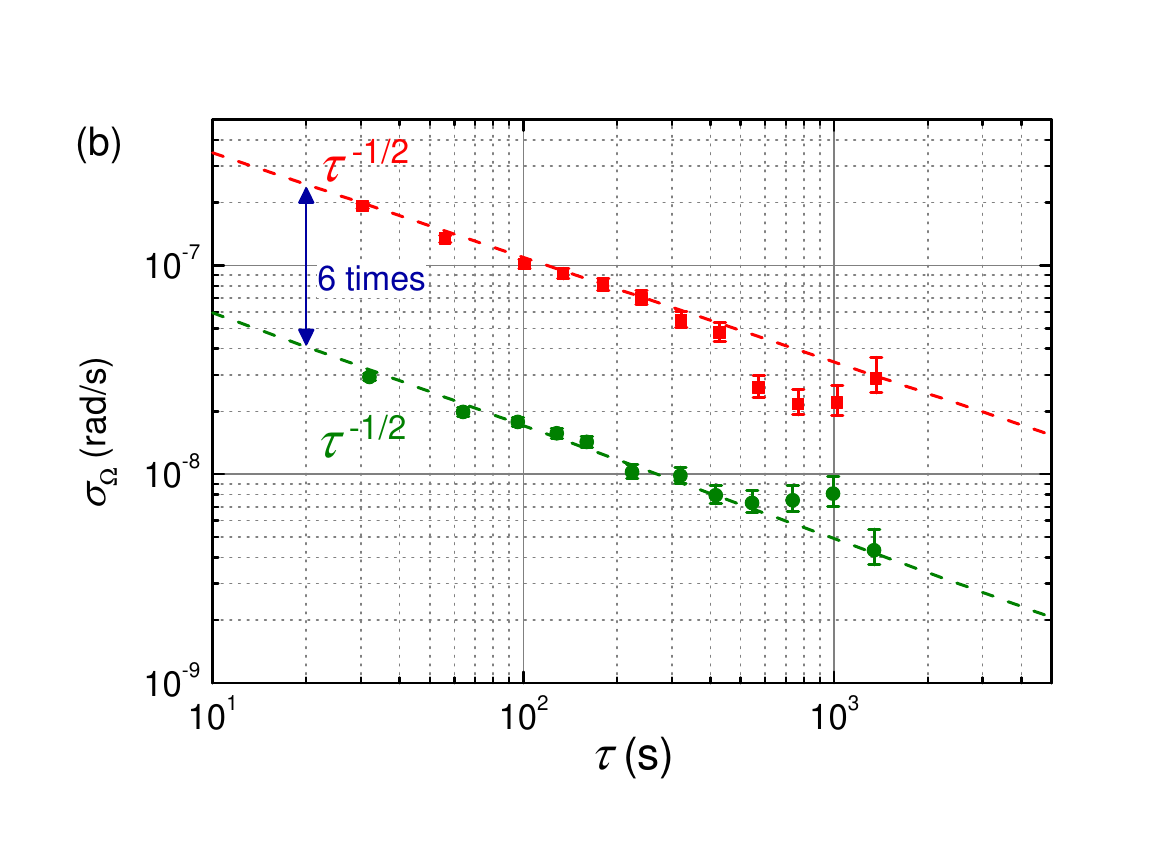}
\end{center}
\caption{\label{fig:GyroVib} (a) Transition probability $P$ vs the
  computed vibration phase $\Phi_{\rm vib}$.  (b) ADEV of the measured
  rotation rate before (squares) and after (circles) correlation.}
\end{figure}

We have also applied the same isolation and correlation techniques to
the LNE-SYRTE gyroscope, which uses cesium-133 atoms in a fountain
configuration and realizes a four-pulse interferometer.  The general
idea was first discussed in~\cite{CanuelSixAxis2006}, and the
experimental setup was described in~\cite{MeunierDuttaStability2014}.
Here, two mechanical accelerometers are used to record the vibration
of the experimental structure.  A weighted sum of these signals yields
the vibration noise $\Phi_{\rm vib}$ at the center of the
interferometer.  Given the large interrogation time of up to $800$~ms
and a corresponding macroscopic Sagnac area of $11$~cm$^2$, $\Phi_{\rm
  vib}$ is often beyond $20$~rad peak-to-peak [see
  Fig.\ \ref{fig:GyroVib}~(a)], rendering a linear correction
impossible.  Instead, we fit $20$ data points to a sinusoid in order
to obtain the rotation phase $\Phi_{\rm sig} = \Phi_\Omega$.  This
gives a vibration rejection efficiency of $6$, as shown in
Fig.\ \ref{fig:GyroVib}~(b).  We thus achieve a short-term stability
of $160$~nrad.s$^{-1}$ at $1$~s and a long-term stability of
$4.4$~nrad.s$^{-1}$ after $1300$~s.

We have seen that passive isolation and post correlation with classical
sensors amount to apply a low pass filter that corrects the atomic sensor from
higher-frequency vibration noise.  In practice, the efficiency of this method
is limited by the intrinsic noise and the nonlinearity of the classical
sensor, and by the fact that the recorded vibration noise does not perfectly
reproduce the vibration of the retroreflection mirror seen by the atoms.
Nevertheless, atomic sensors such as CAG equipped with this system combine
high performance and transportability, and can therefore perform precision
measurements for metrology at different sites~\cite{FarahUnderground2014}.  In
particular, CAG has been participating in international comparisons of
absolute gravimeters since 2009~\cite{Francis2012,Jiang2012}, and its
measurements are in excellent agreement with those from other instruments.

\section{Vibration Compensation in Real Time and Hybridizing Atomic and Classical Sensors}
\label{sec:hybrid}

Environmental stability is rarely guaranteed in field applications.
Moreover, inertial navigation requires the sensors to follow the
moving frame of the vehicle, hence necessitates the removal of any
isolation~\cite{MerletOperating2009,GeigerDetecting2011}.  Such
scenarios, together with the intrinsic high sensitivity of AI-based
inertial sensors, result in the scatter of the data points beyond the
linear regime, and thus a reduction of sensitivity.  We demonstrate a
procedure of compensating the vibration phase in real time, and the
hybridization of an atomic and a classical
accelerometers~\cite{LautierHybridizing2014}.

\begin{figure}[hbt!]
\begin{center}
\includegraphics[height=4cm, trim = 4.5cm 3.5cm 7.8cm 2cm, clip]{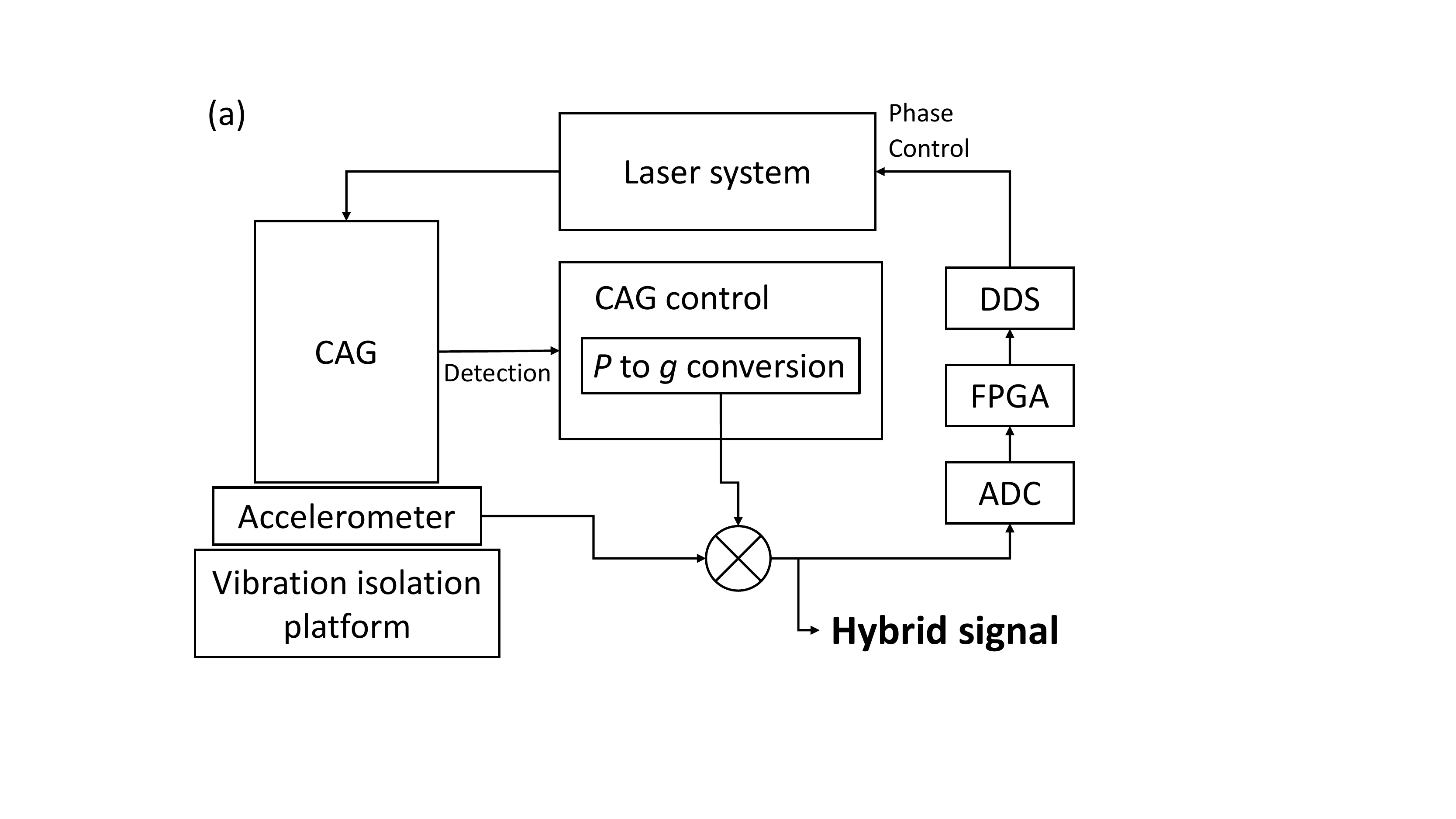}\hspace{.5cm}%
\includegraphics[height=4cm, trim = .6cm .6cm 1cm 1cm, clip]{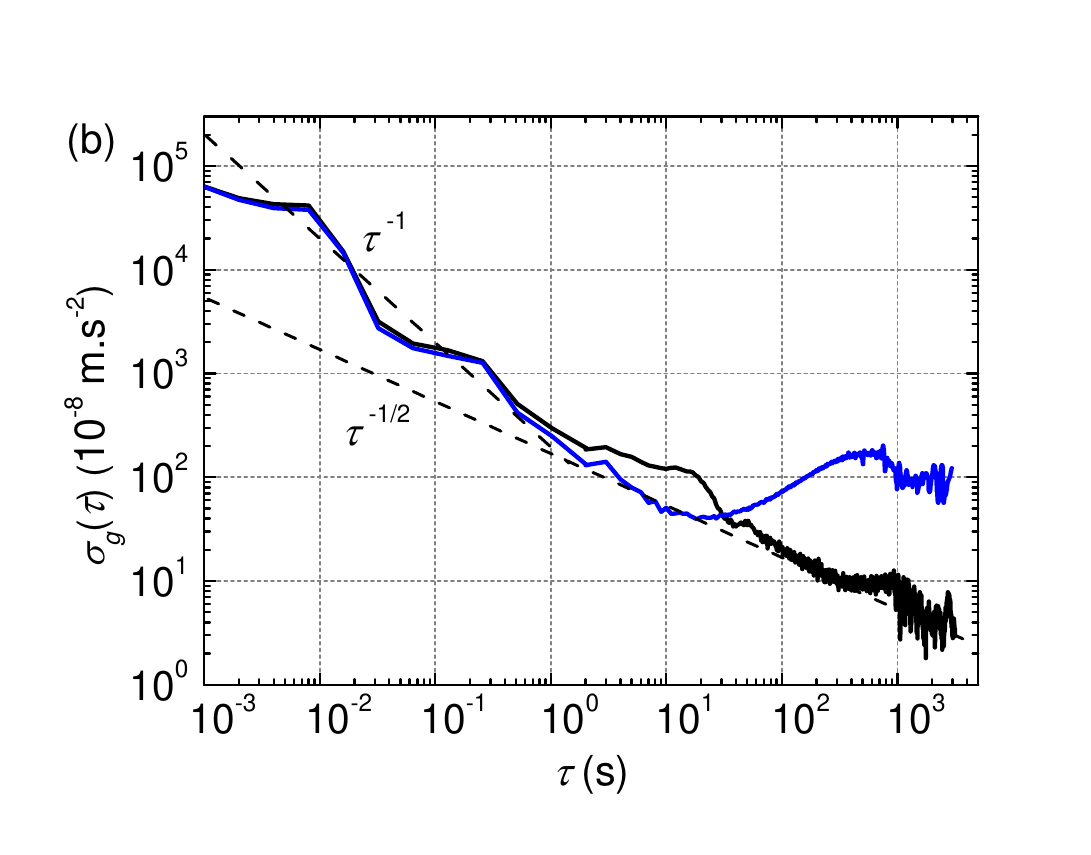}
\end{center}
\caption{\label{fig:hybrid} (a) Illustration of the real-time-compensation
  scheme and the hybridization with classical sensors.  (b) ADEV of a
  classical accelerometer (blue) and a hybrid gravimeter (black).  (b) is
  originally published in~\cite{LautierHybridizing2014}.  }
\end{figure}

We operate a previous prototype of
CAG~\cite{LeGouetLimits2008,MerletSimple2014} in Paris city.  The
vibration signal is acquired using a mechanical accelerometer rigidly
fixed to the experimental frame.  A field programmable gate array
(FPGA) based calculator treats the digitized acceleration signal to
estimate the vibration phase $\Phi_{\rm vib}$.  About $400~\mu$s
before the final interrogation pulse, the FPGA commands a phase jump
in the direct digital synthesizer (DDS) that controls the phase lock
of the Raman lasers.  Such a feed forward scheme [see
  Fig.\ \ref{fig:hybrid}~(a)] allows us to operate the atomic sensor
at mid fringe with little scatter.  The vibration rejection efficiency
can be as high as $60$.

Further more, we perform a hybridization of the
accelerometer with CAG.  The DC component of conventional
accelerometers suffers from drift and temperature dependence that
severely degrades the long-term stability of the acceleration
measurement.  By locking the DC component on the CAG output, we
realize a hybrid sensor which combines the large bandwidth (DC to
$430$~Hz) of the accelerometer and the bias stability and the accuracy
of CAG.  We show in Fig.\ \ref{fig:hybrid} (b) the ADEV of the hybrid
sensor placed on a vibration isolation platform.  The short-term
stability of the sensor is determined by the self-noise of the
accelerometer and vibrations.  The ADEV decreases as $\tau^{-1}$, as
high-frequency vibrations are averaged without dead time by the
accelerometer.  The long-term stability decreases as $\tau^{-1/2}$,
indicating that it is set by CAG operating in the sequential mode and
limited by the quality of the correlation.  The added value of this
technique for inertial navigation is assessed
in~\cite{LautierHybridizing2014}.

\section{Eliminating Dead Time}
\label{sec:joint}

We have seen that the long-term stability of a hybrid inertial
  sensor still suffers from the dead time of the AI operating in sequential
  mode, which prevents efficient averaging of low-frequency noise (typically
  due to vibration).  We present here a demonstration of a continuous atom
interferometer without dead time, which paves the way to the intrinsically
limited performance of AI-based inertial sensors.

\begin{figure}[hbt!]
\begin{center}
{\includegraphics[height=4cm]{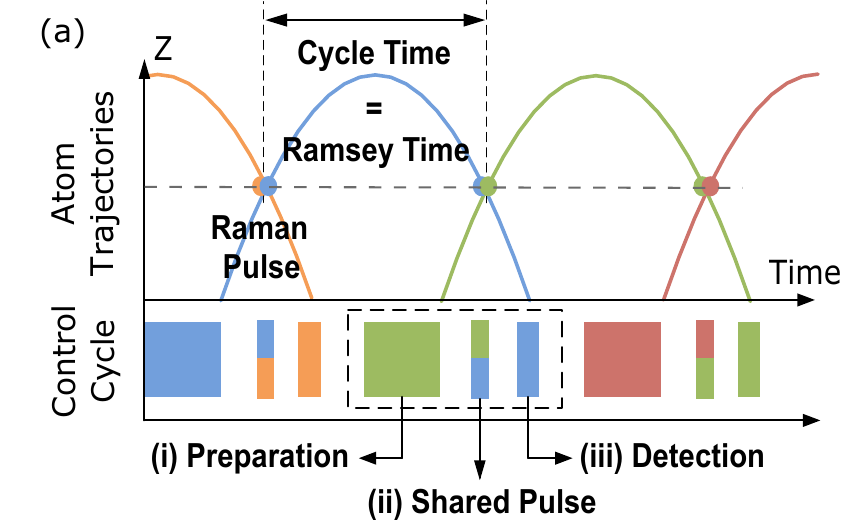}}\hspace{.5cm}%
{\includegraphics[height=4cm]{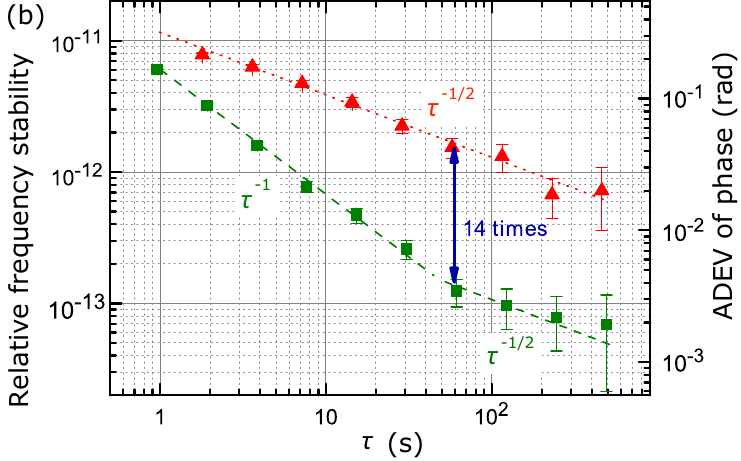}}
\end{center}
\caption{\label{fig:joint} (a) Illustration of the zero-dead-time sequence of
  the atom gyroscope in the clock mode.  (b) ADEV of the atomic clock in
  normal (triangles) and continuous (squares) mode.  This figure is originally
  published in~\cite{MeunierDuttaStability2014}.}
\end{figure}

We operate the fountain gyroscope mentioned in
Sec.\ \ref{sec:correlation} as an atomic
clock~\cite{MeunierDuttaStability2014}, using a sequence of two Raman
pulses ($\pi/2$-$\pi/2$) so that $\Delta \Phi = \phi_1-\phi_2$.  We
add a white noise over a bandwidth of $400$~Hz to the local oscillator
(LO) which controls the phase lock of the two Raman lasers.  The added
noise is analogous to the vibration noise in atomic inertial sensors.
The cycle time and the Ramsey time are matched to be $480$~ms, and
each Raman pulse is shared between two atom clouds, one leaving and
the other entering the interrogation zone, as shown in
Fig.\ \ref{fig:joint} (a).  This imprints the same LO noise on
consecutive interferometric phases , i.e.\ $\phi_2^i = \phi_1^{i+1}$,
allowing cancellation of the LO noise in the interferometric phase
accumulated over multiple shots.  As a result, the ADEV of the
relative frequency stability of the continuous clock follows the
$\tau^{-1}$ scaling, as shown in Figure\ \ref{fig:joint} (b)
(squares).  The same figure shows the ADEV of a sequential clock
(triangles), which follows the $\tau^{-1/2}$ scaling.  The continuous
clock gives a 14-fold gain in the relative frequency stability at
$60$~s.  Thereafter, both are limited by uncorrelated noise
(e.g. detection noise), and follow the $\tau^{-1/2}$ scaling.  Note that the performance of the continuous clock after $60$~s coincides with that of a sequential clock without the additional noise in the LO.  

\section{Conclusions}
\label{sec:conclusion}

We have shown various steps towards exploring the full sensitivity of
AI-based inertial sensors, which currently face the problems
associated with the vibration noise, aliasing and the dead time
between measurements.  These problems would only become more severe
when the sensors leave the environmentally stable
laboratories for field applications.  Moreover, the need to measure a
fluctuating inertial signal (e.g.\ in navigation) renders active or
passive isolation undesirable.  Our efforts to correlate and hybridize
atomic and mechanical sensors i) demonstrate AI-based inertial sensors
with high sensitivity, ii) improve the dynamics beyond the linear
regime, and iii) partially remove the dead time between successive
interrogations.  They represent a major step forward in reducing the
size and the complexity of the sensors, making them suitable for
on-board and field applications, and have stimulated the
commercialization of atomic inertial sensors~\cite{muquans}.  Such
instruments now participate in gravity measurements, and have the
potential to perform inertial navigations.  The ability to fully
eliminate the dead time will eventually lead to a new
generation of QPN limited high-performance inertial sensors exploiting
the full potential of atom interferometry.

\ack

The authors would like to thank the Institut Francilien pour la Recherche sur
les Atomes Froids (IFRAF), the Agence Nationale pour la Recherche
(ANR-9-BLAN-0026-01) in the framework of the MiniAtom collaboration, the
Action Sp\'ecifique Gravitation R\'ef\'erences Astronomie et M\'etrologie
(GRAM), the Centre National d'Etudes Spatiales (CNES), the D\'el\'egation
G\'en\'erale pour l'Armement (DGA), the LABEX Cluster of Excellence FIRST-TF
(ANR-10-LABX-48-01) within the Program ``Investissements d'Avenir'' operated
by ANR, the city of Paris through the Emergence programme (project
HSENS-MWGRAV), and the European Metrology Research Programme (EMRP) within the
kNOW project for financial support.  The EMRP was jointly funded by its
participating countries within the European Association of National Metrology
Institutes (EURAMET) and the European Union.


\end{document}